\documentclass{iaus}
\usepackage{graphicx}
\title[Optimized gyrosynchrotron algorithms and fast codes]{Optimized gyrosynchrotron algorithms and fast codes}
\author[Alexey Kuznetsov \& Gregory Fleishman]{Alexey A. Kuznetsov$^{1,2}$ \and Gregory D. Fleishman$^{3,4}$}
\affiliation{$^1$Armagh Observatory, Armagh BT61 9DG, Northern Ireland, UK\\
email: {\tt aku@arm.ac.uk}\\[\affilskip]
$^2$Institute of Solar-Terrestrial Physics, Irkutsk 664033, Russia\\[\affilskip]
$^3$New Jersey Institute of Technology, Newark, NJ 07102, USA\\
email: {\tt gfleishm@njit.edu}\\[\affilskip]
$^4$Ioffe Physico-Technical Institute, St. Petersburg 194021, Russia}
\pubyear{2010}
\volume{274}
\pagerange{1--3}
\jname{Advances in Plasma Astrophysics}
\editors{A. Bonanno, E. de Gouveia Dal Pino, \&  A. Kosovichev, eds.}
\begin{document}
\maketitle
\begin{abstract}
Gyrosynchrotron (GS) emission of charged particles spiraling in
magnetic fields plays an exceptionally important role in
astrophysics. In particular, this mechanism makes a dominant
contribution to the continuum solar and stellar radio emissions.
However, the available exact equations describing the emission
process are extremely slow computationally, thus limiting the
diagnostic capabilities of radio observations. In this work, we
present approximate GS codes capable of fast calculating the
emission from anisotropic electron distributions. The computation
time is reduced by several orders of magnitude compared with the
exact formulae, while the computation error remains within a few
percent. The codes are implemented as the executable modules
callable from IDL; they are made available for users via web sites.
\end{abstract}

Gyrosynchrotron (GS) radiation (incoherent radiation of mildly
relativistic electrons spiraling in a magnetic field) makes a
dominant contribution into the radio and microwave emission of solar
and stellar flares. Therefore, radio observations potentially can be
used to diagnose the parameters of the emission sources (such as
energy and pitch-angle distributions of the energetic electrons,
magnetic field, and plasma density). The diagnostics can be made,
e.g., by using the forward-fitting methods (e.g., \cite[Fleishman et al.
2009]{fle09}), which requires fast and accurate methods of calculating the
GS emission.

The exact formulae for the gyromagnetic radiation are known for
several decades (\cite[Eidman 1958]{eid58}, \cite[1959]{eid59};
\cite[Melrose 1968]{mel68}; \cite[Ramaty 1969]{ram69}). However, these
formulae are computationally slow, especially when high harmonics of
the cyclotron frequency are involved (i.e., in a relatively weak
magnetic field). A number of simplified approaches has been proposed
(e.g., \cite[Petrosian 1981]{pet81}; \cite[Dulk \& Marsh
1982]{dul82}; \cite[Klein 1987]{kle87}), but those approximations
were developed for the isotropic electron distributions only, while
real distributions in the solar and stellar flares are often
expected to be highly anisotropic. Thus, there is a need to develop
algorithms and computer codes that would be fast, accurate, and
applicable to both isotropic and anisotropic electron distributions.

\begin{figure}
\includegraphics{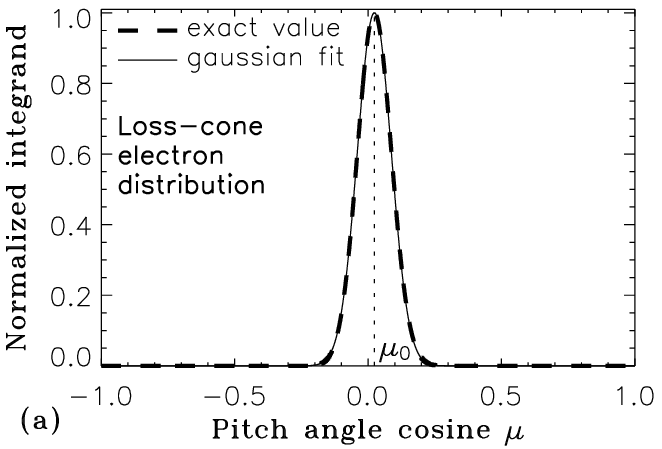}
\includegraphics{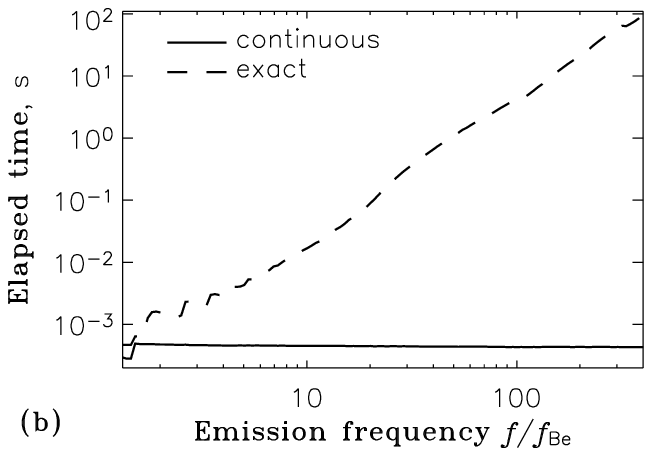}
\caption{(a) Example of the angular integrand (dashed line) and its
gaussian fit (solid line). (b) Time required to calculate the
intensity and polarization of the GS emission at a given frequency
(a 2 GHz Intel Pentium processor was used for the calculations).} 
\label{fig01}
\end{figure}

In developing such algorithms, we follow the approximation proposed
by \cite{pet81} and  \cite{kle87}, which has been proved to provide
a reasonable accuracy in a high-frequency range in the case of the
isotropic or weakly anisotropic electron distributions. We have
found that, after a few improvements (involving a wider use of
numerical methods), this approximation can be extended to the
strongly anisotropic distributions, provided that the pitch-angle
distribution function is continuous together with its first and
second derivatives. The key element of this new algorithm (which we
call ``continuous'') is a possibility to approximate the angular
integrand of the exact GS equations by a Gaussian fit, see
Fig.~\ref{fig01}a, which is then integrated analytically saving
the computation time significantly (see Fig.~\ref{fig01}b). The
comprehensive description of the new approximation 
is given in the recently published article of \cite{fle10}. Besides
a wider applicability range, the new approximation improves the
accuracy for the isotropic distributions in comparison with the
original Petrosian-Klein algorithm.

\begin{figure}
\includegraphics{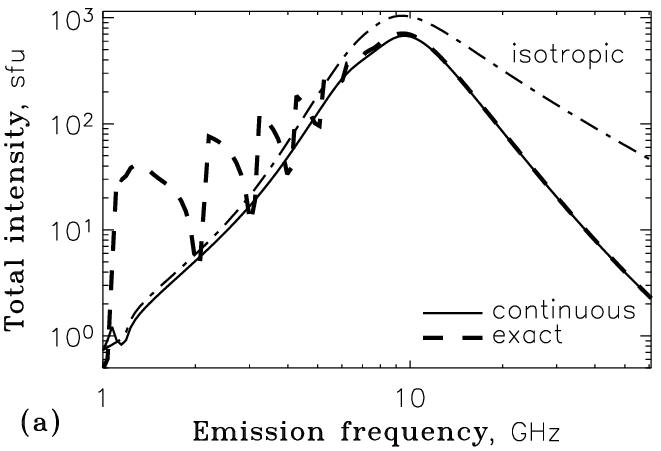}
\includegraphics{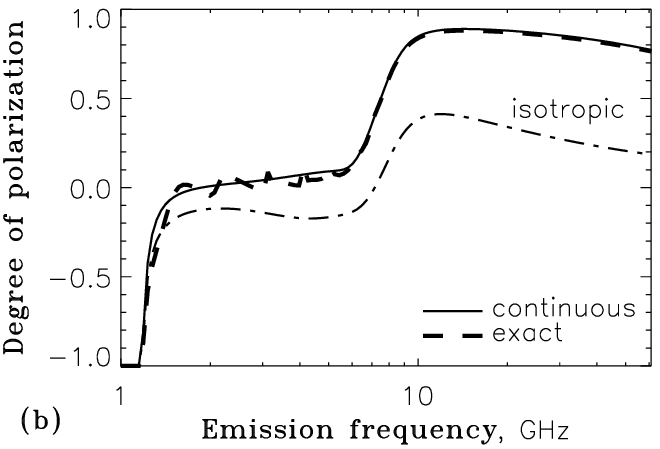}\\
\includegraphics{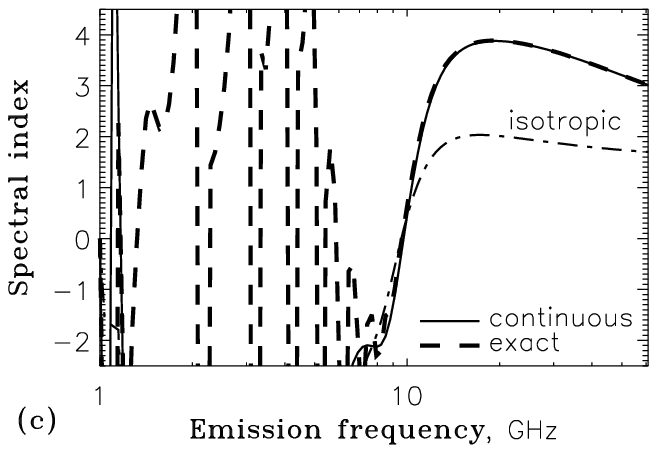}
\includegraphics{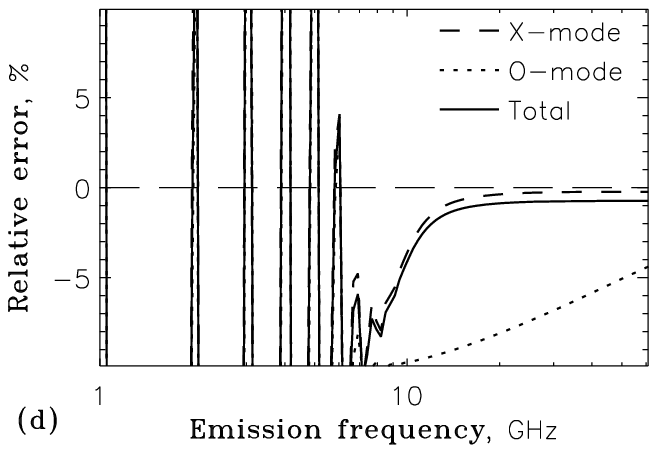}
\caption{Calculated parameters of the GS emission for the electron
distribution of loss-cone type. In the panels (a-c), the dash-dotted
line shows the corresponding parameters for the isotropic
distribution. Panel (d) shows the relative difference between the
emission intensities calculated using the exact and continuous
codes.} \label{fig02}
\end{figure}

Figure \ref{fig02} demonstrates an example of the GS emission
calculated using the exact formulae and new continuous algorithm.
The energetic electrons are assumed to have a power-law distribution
over energy and a loss-cone distribution over pitch-angle. One can
see that the contunuous algorithm provides very high accuracy,
especially at high frequencies. The degree of polarization is
reproduced with high accuracy. Spectral index of the emission in the
optically thin range is reproduced very well, too. At low
frequencies, the continuous algorithm is unable to reproduce the
harmonic structure of the GS emission; however, the mean level of
the spectrum is reproduced well. In Figs. \ref{fig02}a-\ref{fig02}c,
an additional line shows the corresponding parameters calculated for
the isotropic electron distribution. One can see that the considered
anisotropy affects significantly the emission intensity,
polarization, and spectral index. Nevertheless, the continuous
algorithm works excellently even for this (strongly anisotropic)
distribution.

Being that accurate, the continuous algorithm is 
much faster than the exact expressions. As one can notice from Fig.
\ref{fig01}b, computation time for the exact formulae grows
exponentially with the harmonic number, while for the continuous
algorithm this time is nearly constant. As a result, at high
frequencies (i.e., where the continuous algorithm is very accurate),
the computation time can be reduced by several orders of magnitude
in comparison with the exact formulae. In addition to this
continuous algorithm, \cite{fle10} developed a number of codes,
which can be gradually tuned to optimize either computation time or
accuracy including recovery of the low-frequency harmonic structure
of the GS radiation, remaining much faster than the exact codes.

The new algorithms are implemented as executable modules (Windows
dynamic link libraries and Linux shared objects) callable from IDL.
The codes are called using the IDL \verb|call_external| function.
This approach combines high computation speed with the IDL
visualization capabilities. A factorized form of the electron distribution function is adopted:
$F(E, \alpha)=u(E)g(\alpha)$, where $E$ and $\alpha$ are the
electron energy and pitch-angle, respectively. Currently, the code contains 9
built-in energy distributions (including thermal, power-law, kappa,
etc.) and 5 pitch-angle distributions (including isotropic,
loss-cone, and beam-like); any combination of the energy and
pitch-angle distribution is possible. The code returns the
intensities of the ordinary and extraordinary modes from a
homogeneous source located at the Sun (and observed at the Earth),
as well as the corresponding gyroabsorption factors which can
directly be used for solving the equation of radiation transfer in
an inhomogeneous case.
Fast GS codes together with the comprehensive documentation are
available as the online supplement to the article of \cite{fle10},
and at the web site: \verb|star.arm.ac.uk/~aku/gs/|

\begin{acknowledgements}
A.K. thanks the Leverhulme Trust for financial support. This work
was supported in part by NSF grants ATM-0707319, AST-0908344, and
AGS-0961867 and NASA grant NNX10AF27G to New Jersey Institute of
Technology, and by the RFBR grants 08-02-92204, 08-02-92228, 09-02-00226, and
09-02-00624. 
\end{acknowledgements}

\end{document}